\documentclass{article}
\usepackage{ijcai23}

\usepackage{times}
\usepackage{soul}
\usepackage{url}
\usepackage[hidelinks]{hyperref}
\usepackage[utf8]{inputenc}
\usepackage[small]{caption}
\usepackage{graphicx}
\usepackage{amsmath}
\usepackage{amsthm}
\usepackage{booktabs}
\usepackage{algorithm}
\usepackage{algorithmic}
\usepackage[switch]{lineno}

\usepackage{amsfonts}
\usepackage{mathtools}
\usepackage{subcaption}
\usepackage{musicography}

\newtheorem{example}{Example}
\newtheorem{theorem}{Theorem}

\DeclarePairedDelimiterX{\infdivx}[2]{(}{)}{%
  #1\;\delimsize\|\;#2%
}
\newcommand{\infdiv}{{\mathbb K}{
\mathbb L}\infdivx}

\title{Q\&A: Query-Based Representation Learning for Multi-Track Symbolic\\Music re-Arrangement}

\author{
Jingwei Zhao$^{2,3}$\and
Gus Xia$^{4,5}$\And
Ye Wang$^{1,2,3}$
\affiliations
$^1$School of Computing, NUS\\
$^2$Institute of Data Science, NUS\\
$^3$Integrative Sciences and Engineering Programme, NUS Graduate School\\
$^4$Music X Lab, NYU Shanghai\\
$^5$MBZUAI\\
\emails
\href{mailto:jzhao@u.nus.edu}{jzhao@u.nus.edu},
\href{mailto:gxia@nyu.edu}{gxia@nyu.edu},
\href{mailto:wangye@comp.nus.edu.sg}{wangye@comp.nus.edu.sg}
}

\begin{document}

\maketitle

\begin{abstract}
 
Music rearrangement is a common music practice of reconstructing and reconceptualizing a piece using new composition or instrumentation styles, which is also an important task of automatic music generation. Existing studies typically model the mapping from a source piece to a target piece via supervised learning. In this paper, we tackle rearrangement problems via self-supervised learning, in which the mapping styles can be regarded as conditions and controlled in a flexible way. Specifically, we are inspired by the representation disentanglement idea and propose Q\&A, a query-based algorithm  for multi-track music rearrangement under an encoder-decoder framework. Q\&A learns both a \emph{content} representation from the mixture and function (\emph{style}) representations from each individual track, while the latter queries the former in order to rearrange a new piece. Our current model focuses on popular music and provides a controllable pathway to four scenarios: 1) re-instrumentation, 2) piano cover generation, 3) orchestration, and 4) voice separation. Experiments show that our query system achieves high-quality rearrangement results with delicate multi-track structures, significantly outperforming the baselines.
\end{abstract}

\section{Introduction}
It is sometimes easy to craft an idea of the melody but usually hard to frame a good arrangement. Formally, \textit{arrangement} refers to the form of a musical piece, typically with textures and voicing carefully designed for multiple instruments as a unique style. On top of that, a piece can also be rearranged to convey new feelings. Such \emph{rearrangement} scenarios include piano cover generation from multi-track music, multi-track orchestration from piano, and re-instrumentation using varied instruments, which are all common tasks in music practice. 


While much progress has been witnessed in automatic music generation \cite{huang2018music,huang2020pop,hsiao2021compound},  rearrangement remains a challenging problem. Among various ways to rearrange a piece, most studies have focused on the reduction from complex forms to simpler ones, such as generating piano covers from multi-track music. The reduction is typically done by masking least significant notes, either identified by rule-based criteria \cite{nakamura2018statistical} or learned by supervision \cite{terao2022difficulty}. While rearrangement in this way is generally faithful, it tends to produce sparse or repetitive textures that fall short of creativity. More recent works have also taken on simple-to-complex rearrangement, such as orchestration \cite{crestel2016live,dong2021towards}. While these works are still fully supervised, the scarce of paired piano and multi-track data remains a major problem in this direction.

Another considerable challenge for music rearrangement lies in \emph{multi-track} modelling. Previous works have typically interpreted ``track'' as ``instrument'' and merge individual tracks of the same instrument class to simplify the problem \cite{dong2018musegan,ren2020popmag}. However, instrument alone is not necessarily a good representative of a multi-track system in symbolic music. For example, pop music often has two guitar tracks of quite different functions -- a \emph{melodic} one and a \emph{harmonic} one. When merged together, the distinctive texture structures of each track become less transparent, which may add extra burden to the model.


\begin{figure*}
 \centerline{
 \includegraphics[width=1.98\columnwidth]{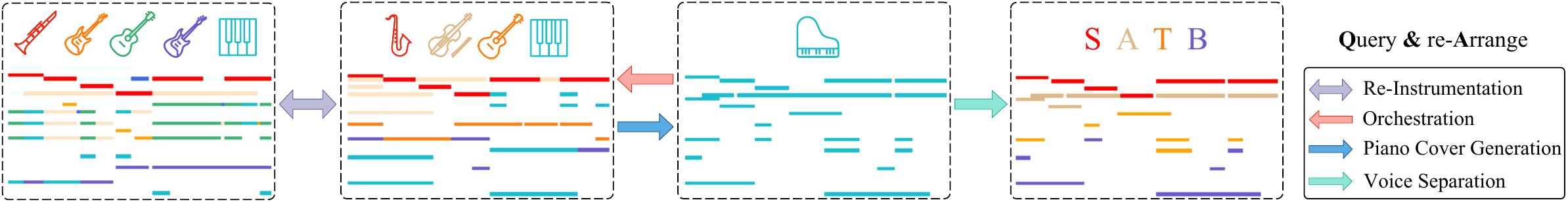}}
 \caption{Q\&A is a unified framework for re-instrumentation, orchestration, piano cover generation, and voice separation.}
 \label{unified_tasks}
\end{figure*}

In this paper, we aim to approach multi-track music rearrangement while balancing faithfulness with creativity. We render the content of a source piece using the style from a reference piece that is free to choose. In terms of the ``style'' of a multi-track piece, apart from instruments, we believe the \emph{function} of each component track is also important. Specifically, we consider the function of a track as 
 its texture density distribution along the time- and pitch-axes, respectively, which can describe both the distinctive intra-track structures (\textit{e.g.}, melodic \textit{v.s.} harmonic) and the inter-track dependencies (\textit{e.g.}, pitch range and voicing). We use track functions as queries in a \emph{query-based track separation process} to reconstruct individual tracks from a track-wise condensed mixture. Under the VAE framework \cite{wang2020pianotree}, we devise a pipeline consisting of four components: 1) an encoder that maps a mixture to the latent space; 2) a query-net \cite{lee2019audio} that encodes function features of each track; 3) a Transformer-based \cite{vaswani2017attention} query system that separates each track from the mixture at the latent representation level; and 4) a decoder that reconstructs each separated track. At inference time, our model can query a piece and rearrange it with diverse track functions. 

We name our model after \emph{Q\&A} (Query \& re-Arrange), which provides a unified solution to a range of multi-track music rearrangement tasks, including: 1) \emph{re-instrumentation} -- to rearrange a multi-track piece with a new track system; 2) \emph{piano cover generation} -- to rearrange a multi-track piece into piano solo; and 3) \emph{orchestration} -- to rearrange a piano piece with a variable types of instruments in a variable number of tracks. By inferring track functions as voice hints, our model can additionally handle 4) \emph{voice separation} -- to separate distinctive voicing tracks (assuming a preset total number of voices) from an ensemble mixture by generating each track. Figure \ref{unified_tasks} shows the relations among the four tasks.

Our current model focuses on pop music rearrangement. We also test our model's voice separation performance on string quartets and Bach chorales. Experimental results show that our model not only generates high-quality arrangements, but also maintains fine-grained symbolic track structures with musically intuitive and playable textures for each track. In summary, our contributions in this paper are as follows:

\begin{itemize}
    \item \textbf{A versatile rearrangement model}: We present \emph{Q\&A}\footnote{\href{https://github.com/zhaojw1998/Query-and-reArrange}{https://github.com/zhaojw1998/Query-and-reArrange}}, the first unified framework for re-instrumentation, piano cover generation, orchestration, and voice separation. The rearrangement results demonstrate state-of-the-art quality over existing models for similar purposes.

    \item \textbf{Function-aware multi-track music modelling}: We design instrument-agnostic track functions for multi-track modelling, which can better describe the distinctions of parallel tracks and their dependencies. This method is applicable to a wider range of music generation tasks.

    \item \textbf{Query-based representation learning}: We introduce a self-supervised query system separating parts from the mixture at latent representation level. Experiments show that our model learns style representations of each part disentangled from the mixture content, demonstrating interpretable and controllable generative modelling.
\end{itemize}

\section{Related Work}

\subsection{Symbolic Music Rearrangement}

Existing studies on music rearrangement commonly rely on supervised learning to map a source piece to a target one. For example, Crestel and Esling \shortcite{crestel2016live} project piano solo to orchestra by training a seq2seq model on a classical repertoire of paired data \cite{DBLP:conf/ismir/CrestelEHM17}. Dong \textit{et al.} \shortcite{dong2021towards} approach automatic instrumentation by predicting the instrument attribute of each note in a track-wise condensed mixture. Models for piano reduction can have more rule-based designs \cite{nakamura2015automatic,takamori2017automatic}, but are still generally under supervised frameworks. Except for several works that consider difficulty level as condition \cite{nakamura2018statistical,terao2022difficulty}, most models cannot steer the rearrangement process or change the composition style.


In this paper, instead of supervised mapping, we render the content of a source piece using the style from a reference piece. In terms of content, we preserve the general melodic and harmonic structures. As for style, we introduce a new track system, \textit{i.e.}, textural functions of each track along with the instruments to play them, to reconceptualize the source piece. Our methodology can be formalized as composition style transfer \cite{dai2018music} while existing research most relevant to us is \cite{hung2019musical}, which approaches re-instrumentation by transferring instrument timbres from different references. While Hung \textit{et al.} \shortcite{hung2019musical} still require supervision from audio-symbolic pairs, our model is fully symbolic-based, self-supervised, and unified for re-instrumentation, piano cover generation, and orchestration.  

\begin{figure*}
 \centerline{
 \includegraphics[width=2\columnwidth]{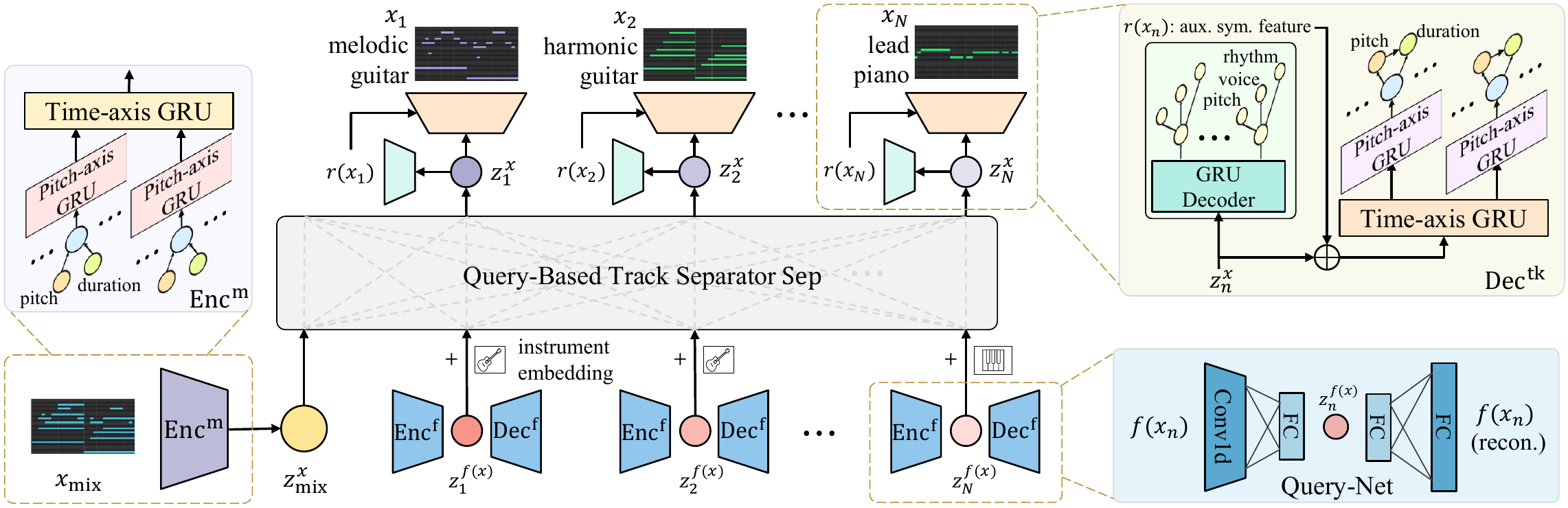}}
 \caption{The architecture of Q\&A consists of four key components: mixture encoder $\mathrm{Enc}^\mathrm{m}$, function query-net with encoder $\mathrm{Enc}^\mathrm{f}$ and decoder $\mathrm{Dec}^\mathrm{f}$, track separator $\mathrm{Sep}$, and track decoder $\mathrm{Dec}^\mathrm{tk}$. Q\&A learns both a content representation from the mixture and function representations from each individual track, while the latter queries the former in order to rearrange a new piece.}
 \label{model}
\end{figure*}

\subsection{Multi-Track Music Modelling}
Multi-track music is an arrangement form commonly seen in accompaniment, symphony, ensembles, \textit{etc.} However, it is very challenging for machines to understand multi-track data. To capture inter-track dependency, mainstream approaches either distribute vari-instrument tracks into parallel data channels \cite{dong2018musegan,zhu2018xiaoice,hung2019musical} or incorporate instrument labels as part of note event tokens \cite{donahue2019lakhnes,musenet,ren2020popmag}. Such methods are ideal for CNN-based and language models, respectively, yet both inevitably merging co-instrument tracks together. The event-based approach additionally enforces a positional relation to parallel tracks that are not sequentially ordered, which can damage the intrinsic structure of multi-track music \cite{wang2021musebert}. More recently, several works target at these issues and support generating co-instrument tracks without a sequential assumption \cite{ens2020mmm,liu2022symphony}. However, these models are not applicable to general music rearrangement.

In this work, apart from instrument, we introduce track function as an equally (if not more) important feature to describe and distinguish individual tracks in multi-track music. We define and use the function of a track to represent its texture and voicing structures. We further model multi-track music via self-supervised learning, \textit{i.e.}, using each function to query and separate corresponding tracks from a mixture. 


\section{Methodology}
We propose Q\&A, a query-based algorithm for multi-track music rearrangement under an encoder-decoder framework. An overview of our model is shown in Figure \ref{model}. In this section, we first introduce our data representation of multi-track music and track functions in Section \ref{data_represent}. Then, we introduce our model architecture and training objectives in Section \ref{model_architect} and \ref{train_objective}. Finally, in Section \ref{inference}, we elaborate on how Q\&A can be applied to music rearrangement at inference time.

\subsection{Data Representation}\label{data_represent}
\subsubsection{Multi-Track Music}
 Given multi-track music $x$ with $N$ tracks, our model aims to reconstruct each track $x_n$, where $n=1, 2, \cdots, N$, from a track-wise condensed mixture $x_\mathrm{mix}$. We represent $x_n$ in the modified piano-roll format proposed by \cite{wang2020learning}, and $x$ as an $N$-track collection. Formally,
\begin{equation}
    x = x_{1:N} \coloneqq \{x_n\}_{n=1}^N,
\end{equation}
where individual track $x_n$ is a $P\times T$ matrix. $P=128$ represents 128 MIDI pitches, and $T$ is the time dimension. Each data entry $(p, t)$ of $x_n$ is an integer value representing note duration on the onset positions. The condensed mixture $x_\mathrm{mix}$ is also a $P\times T$ matrix where each entry is the position-wise maximum value across $N$ tracks. In this paper, we consider 2-bar (8-beat) music data segments in $\frac{4}{4}$ time signature quantized at $\frac{1}{4}$ beat unit, deriving $T=32$ time steps for each music sample. We also focus on composition-level aspects while disregarding performance-level dynamics like MIDI velocity. 

\subsubsection{Track Function}
We define the function of each track $x_n$ as its texture density features computed from the modified piano-roll format. Specifically, we define descriptors of pitch function $f^\mathrm{p}(\cdot)$ and time function $f^\mathrm{t}(\cdot)$ as follows:
\begin{align}
    f^\mathrm{p}(x_n) &= \mathrm{rowsum}(\mathbf{1}_{\{x_n > 0\}}) / T,\\
    f^\mathrm{t}(x_n) &= \mathrm{colsum}(\mathbf{1}_{\{x_n > 0\}}) / P,
\end{align}
where $\mathbf{1}_{\{\cdot\}}$ is the indicator function expressing individual note onset entries as 1. $\mathrm{rowsum}(\cdot)$ and $\mathrm{colsum}(\cdot)$ each sums up one dimension, resulting in a $P$-D and $T$-D vector, respectively. $f^\mathrm{p}(x_n)$ is essentially a pitch histogram, which is related to key, chord, and the pitch range of $x_n$. $f^\mathrm{t}(x_n)$ indicates voice densities of $x_n$ and is related to rhythmic patterns and grooves. Each vector is normalized to $[0, 1]$. 

\subsection{Model Architecture}\label{model_architect}
Figure \ref{model} shows the overall architecture of our model consisting of four key components: 1) a mixture encoder, 2) a function query-net, 3) a track separator, and 4) a track decoder.

\subsubsection{Mixture Encoder}
As $x_\mathrm{mix}$ is a single-track polyphony, we use the encoder module of PianoTree VAE \cite{wang2020pianotree}, the state-of-the-art polyphonic representation learning model, to encode a 256-D mixture representation $z_\mathrm{mix}^{x}$. The PianoTree encoder first converts $x_\mathrm{mix}$ to a compact and ordered note event format, where each event contains pitch and duration attributes. It then applies a pitch-wise bi-directional GRU to summarize concurrent notes at time step $t$ to an intermediate representation $\mathrm{simu\_note}_t$. On top of $\mathrm{simu\_note}_{1:T}$, it further applies a time-wise GRU to encode the full mixture representation $z_\mathrm{mix}^{x}$. The encoding process of PianoTree VAE reflects hierarchical musical understanding from note via chord to grouping, which is interpretable and has proved beneficial for a range of downstream generation tasks \cite{yi2022accomontage2,DBLP:conf/ismir/WuerkaixiBDZ21,zhao2022domain,wang2022audio}.

\subsubsection{Function Query-Net}
The function query-net consists of two VAEs that encode 128-D representations ${z^{\mathrm{p}(x)}_{n}}$ and ${z^{\mathrm{t}(x)}_{n}}$ for track functions $f^\mathrm{p}(x_n)$ and $f^\mathrm{t}(x_n)$, respectively. The pitch and time function encoders each consist of a 1-D convolutional layer with kernel size 12 and 4, respectively. Both are followed by ReLU activation \cite{nair2010rectified} and 1-D max-pooling with kernel size 4 and stride 4. The decoders consist of two fully-connected layers with ReLU activation in between. 
 
It is noted that, with the encoder design, we leverage the translation invariance property of convolution and the blurry effect of pooling \cite{krizhevsky2017imagenet} to discourage the separator from simply retrieving notes that are implied in the track functions. By doing so, our model learns a general style representation instead of the exact density values from the track function. Similar method is also adopted in other VAE architectures to realize disentanglement \cite{wang2020learning}.

\subsubsection{Track Separator}
The track separator is a 2-layer Transformer encoder with 8 attention heads, 0.1 dropout ratio, and GELU activation \cite{hendrycks2016gaussian}. The hidden dimensions of self-attention $d_\mathrm{model}$ and feed-forward layers $d_\mathrm{ff}$ are 512 and 1024, respectively. The input to the separator is a sequence of $N+1$ latent codes including mixture $z_\mathrm{mix}^{x}$ and track functions $z^{f(x)}_{1:N}$, where $z^{f(x)}_n$ denotes the concatenation $[z^{\mathrm{p}(x)}_n; z^{\mathrm{t}(x)}_n]$ as a unified track function representation. We also add learnable instrument embeddings to the corresponding tracks. It is noted that Transformer is permutation-invariant to the index of track functions so that no sequential assumption is enforced. While the self-attention mechanism allows each track function as query to attend to the mixture, it also encourages queries to attend to each other for inter-track dependency. We denote the output of the Transformer as $z^{x}_{1:N}$, which are the expected latent representations for individual tracks $x_{1:N}$. 

\subsubsection{Track Decoder}
We use the decoder module of PianoTree VAE to reconstruct each track $x_n$ from representation $z^{x}_n$. The decoder involves time- and pitch-wise uni-directional GRUs, which mirror the structure of the encoder. To better distinguish parallel tracks, we additionally provide the decoder with an auxiliary time sequence of symbolic features, which are priorly predicted from $z^{x}_n$. Specifically, we consider three auxiliary features: \emph{pitch centre}, \emph{voice intensity}, and \emph{rhythm}, which can serve as strong hints to determine if one track has \emph{melodic}, \emph{harmonic}, and \emph{static} properties \cite{couturier2022annotating,couturier2022dataset}. Both pitch centre and voice intensity are time sequences of scalar values, which indicate centre pitch curve and voice number progression of a track, both normalized to $[0, 1]$. The rhythm feature is a time sequence of onset probabilities, which represents the rhythmic pattern in time. We use a uni-directional GRU to predict the symbolic features from $z^{x}_n$ and feed them to the corresponding time steps of the time-wise GRU in the PianoTree Decoder. Similar method is also applied for disentanglement and reconstruction in \cite{yang2019deep,wang2022audio}. 

\subsection{Training Objectives}\label{train_objective}
The loss terms in our model include 1) reconstruction loss for each track, track functions, and auxiliary symbolic features, and 2) KL loss between all latent representations and standard normal distribution. Our model is essentially a variational autoencoder since the loss function can be formalized as the evidence lower bound (ELBO) of distribution $p(x)$, where $x=x_{1:N}$ is the multi-track music.

The posterior distribution of the VAE is defined as the product of three modules including mixture encoder, query-net encoder, and track separator:
\begin{align}\label{posterior}
q_{\boldsymbol{\phi}}(\mathbf{z}\mid x) :=\; & q_{\phi_1}(z^x_\mathrm{mix} \mid  x_\mathrm{mix}) \prod_{n=1}^N q_{\phi_2}(z^{f(x)}_n \mid  f(x_n)) \nonumber \\
   & \prod_{n=1}^N q_{\phi_3}(z^x_{n} \mid  z^x_\mathrm{mix}, z^{f(x)}_{1:N}),
\end{align}%
where ${\boldsymbol{\phi}} := [\phi_1,\, \phi_2,\, \phi_3]$ denotes the parameters of the three modules, and $\mathbf{z} := [z^x_\mathrm{mix},\, z^x_{1:N},\, z^{f(x)}_{1:N}]$. In Equation \eqref{posterior}, we collectively express two types of track functions as $f(x_n)$ for conciseness. It is noted that both $x_\mathrm{mix}$ and $f(x_n)$ are deterministically transformed from $x$ and hence are not explicitly written in the left-hand side of Equation \eqref{posterior}. 

The reconstruction distribution is defined as the product of three reconstruction terms of query-net decoder, symbolic feature decoder, and track decoder:
\begin{align}
p_{\boldsymbol{\theta}}(x\mid \mathbf{z}) :=
   &\prod_{n=1}^N p_{\theta_1}(f(x_n)\mid z^{f(x)}_n)
    \prod_{n=1}^N p_{\theta_2}(r(x_n)\mid z^x_{n}) \nonumber \\
    &\prod_{n=1}^N p_{\theta_3}(x_n \mid  z^x_{n}, r(x_n)),
\end{align}
where ${\boldsymbol{\theta}} := [\theta_1,\, \theta_2,\, \theta_3]$ denotes the parameters of the three decoders. $r(x_n)$ denotes the auxiliary symbolic features for track $x_n$. The $p_{\theta_1}$ term can be interpreted as a regularizer to the overall output distribution $p_{\boldsymbol{\theta}}(x\mid \mathbf{z})$.

Finally, the overall loss function is as follows:
\begin{align}\label{overall}
    \mathcal{L}({\boldsymbol{\theta}}, {\boldsymbol{\phi}}; x) &= - \mathbb{E}_{\mathbf{z} \sim q_{\boldsymbol{\phi}}} \log p_{\boldsymbol{\theta}}(x \mid \mathbf{z}) \nonumber \\
    &+ \beta \: \infdiv{q_{\boldsymbol{\phi}}(\mathbf{z}\mid x)}{\mathcal{N}(\boldsymbol{0},\,\boldsymbol{1})},
\end{align}
where $\beta$ is a balancing parameter \cite{higgins2016beta}.

\subsection{Style Transfer}\label{inference}
At inference time,  Q\&A can rearrange a multi-track source piece $x=x_{1:N}$ using the track system (style) from a reference piece $y=y_{1:M}$, which can be freely selected. Let $\mathrm{Enc}^\mathrm{m}$, $\mathrm{Enc}^\mathrm{f}$, $\mathrm{Sep}$, and $\mathrm{Dec}^\mathrm{tk}$ be the mixture encoder, function encoder, track separator, and track decoder, respectively, the rearrangement process takes a pipeline as follows:
\begin{align}
    z^x_{\mathrm{mix}} &= \mathrm{Enc}^\mathrm{m}(x_\mathrm{mix}), \nonumber\\
    z^{f(y)}_m &= \mathrm{Enc}^\mathrm{f}(f(y_m)),\; m=1, 2, \cdots, M, \nonumber\\
    z^{x^\prime}_{1:M} &= \mathrm{Sep}(z^x_{\mathrm{mix}},\; z^{f(y)}_{1:M}), \nonumber \\
    x^\prime_m &= \mathrm{Dec}^\mathrm{tk}(z^{x^\prime}_m),\; m=1, 2, \cdots, M,
\end{align}
where $x^\prime=x^\prime_{1:M}$ is the rearrangement result. $x^\prime$ inherits the general harmonic structures from $x$, while also introducing $y$'s track system with new textures, grooves, and track voicing played by a different set of instruments.

In addition to manual selection, reference $y$ can be automatically searched from a database ${\mathcal{D}}$. To guarantee faithful and natural rearrangement results, we develop a simple heuristic to sample $y$ that is ``matched'' with $x$ as follows:
\begin{equation}\label{search}
y = \underset{y \in {\mathcal{D}} }{\mathrm{argmax}} [ \cos(f(y_\mathrm{mix}), f(x_\mathrm{mix})) + \alpha \cdot \epsilon_y ],
\end{equation}
where $\cos(\cdot,\,\cdot)$ measures cosine similarity between the functions (essentially, texture densities) of mixture $y_\mathrm{mix}$ and $x_\mathrm{mix}$, $\epsilon_y \sim {\mathcal{N}}(0,\,1)$ is a noise term for balancing with generality, and $\alpha$ is a balancing parameter. In cases when $y$ and $x$ are very dissimilar, our model robustly follows $x$'s harmony and $y$'s texture and voicing in a general sense of style transfer.


\section{Experiments}

\subsection{Dataset}\label{dataset}
Our model is trained on Slakh2100 \cite{manilow2019cutting} and POP909 \cite{pop909-ismir2020} datasets. In specific, Slakh2100 contains 2K MIDI files of multi-track music, most of which are in pop style. Instruments in Slakh2100 are categorized into 34 classes (while co-instrument tracks are not merged) and each piece contains at least one track of piano, guitar, bass, and drum. In our experiment, we discard the drum track because it does not follow the standard 128-pitch protocol used in other tracks. POP909 is a dataset of 1K pop songs in piano arrangement created by professional musicians. Each piece consists of three piano tracks for vocal melody, lead instrument melody, and piano accompaniment, respectively. By jointly training our model on both datasets, our model can rearrange multi-track music to piano and vice versa. 

\subsection{Training}\label{train_finetune}
For training Q\&A, we use the official training split of Slakh2100 while randomly splitting POP909 (at song level) into training, validation, and test sets with a ratio of 8:1:1. We further augment training data by transposing  each piece to all 12 keys. Our model comprises 19M learnable parameters and is trained with a mini-batch of 64 2-bar segments for 30 epochs on an RTX A5000 GPU with 24GB memory. We use Adam optimizer \cite{kingma2014adam} with a learning rate from 1e-3 exponentially decayed to 1e-5. We apply teacher forcing \cite{toomarian1992learning} for the decoder GRUs in PianoTree VAE with a rate from 0.8 to 0. For the parameter $\beta$ in Equation \eqref{overall}, we apply KL annealing following \cite{wang2022audio} and set $\beta$ increasing from 0 to 0.5 for $z^{f(x)}_n$ and from 0 to 0.01 for the other two factors.

\subsection{Rearrangement Showcase}
\begin{figure*}[t]
 \centerline{
 \includegraphics[width=2\columnwidth]{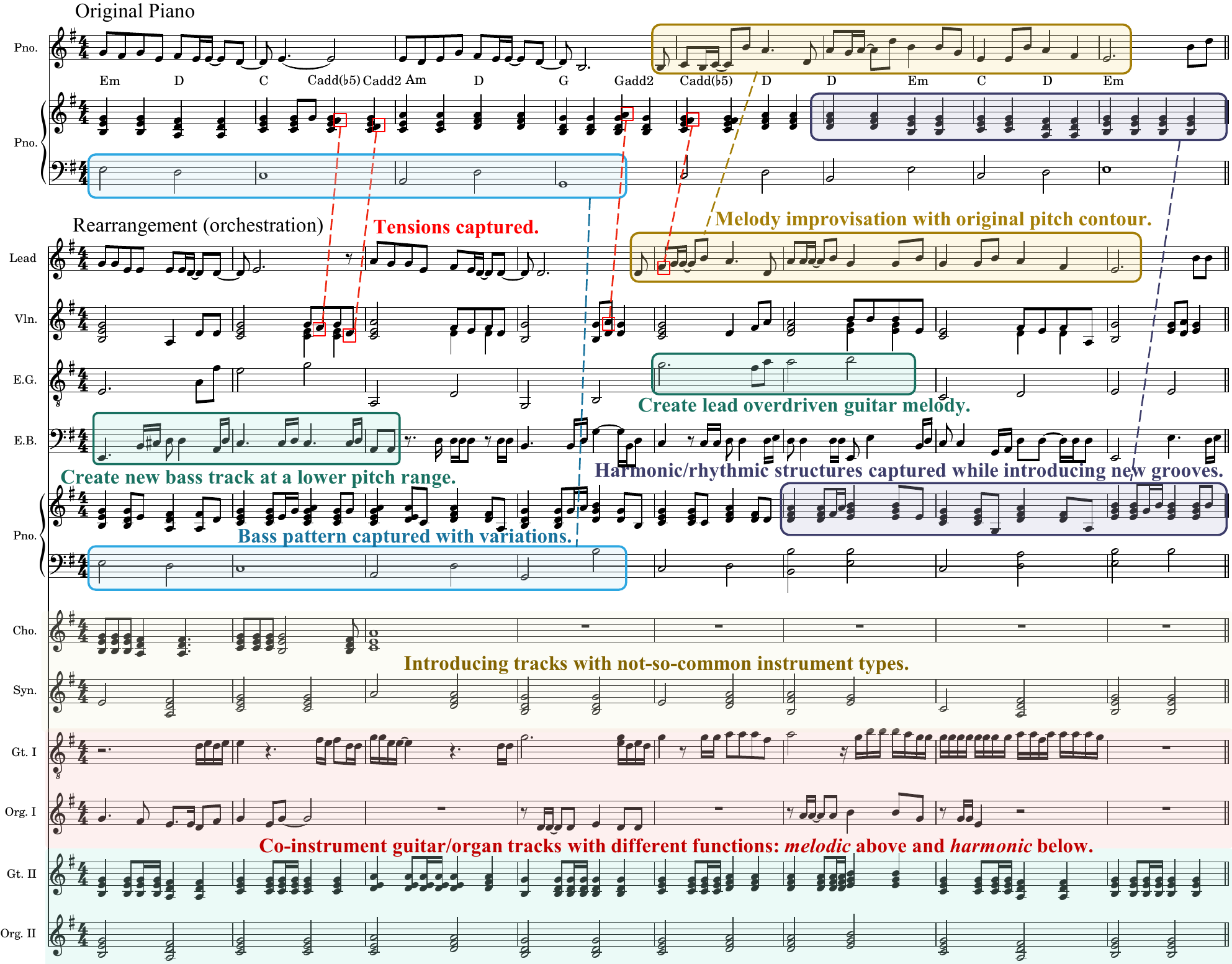}}
 \caption{An orchestration example for {\tt song\_283} from POP909 by our proposed Q\&A model. The result has 11 tracks with varied instruments. Annotations illustrate that the rearrangement is both faithful and creative with a delicate multi-track structure.}
 \label{showcase}
\end{figure*}

An 8-bar rearrangement example (by processing every 2 bars independently) is shown in Figure \ref{showcase}. In specific, this is an orchestration example, where we use Q\&A to rearrange a piano piece into multi-track music. The piano source $x$ is from POP909 while we sample reference $y$ of the same length from Slakh2100 following Equation \eqref{search} with $\alpha=0.2$. We add $f(x_\mathrm{mel})$, the function of $x$'s melody track, to $y$'s track functions as an additional query to guarantee the preservation of the theme melody. Meanwhile, we conduct posterior sampling over $z^{x^\prime}_{\mathrm{mel}}$ to encourage melody improvisation.

In this example, our model rearranges the piano piece into 11 tracks with coherent and delicate multi-track textures. Among the 11 tracks, guitar and organ are each used twice for melodic and harmonic purposes, respectively. Our model preserves the original harmony quite faithfully. Particularly, it captures the added chord notes in $x$ (highlighted by red dotted lines in Figure \ref{showcase}) and retains the tension from the original piece. At the same time, it introduces new groove patterns, bass lines, lead instrument melodies, and a theme melody variation to reconceptualize the piece with more creativity. 



\subsection{Subjective Evaluation on Rearrangement}\label{subjective_rearrangement}
Based on composition style transfer, Q\&A is a unified solution for a range of music rearrangement tasks. In this paper, we focus on orchestration, piano cover generation, and re-instrumentation.  For evaluation, we introduce three existing models as baselines for each of the three tasks as follows: 

\textbf{\textit{BL-Orch.}}: We introduce \textit{Arranger} by \cite{dong2021towards} as our baseline for the orchestration task. We select the BiLSTM variant pre-trained on Lakh MIDI dataset \cite{Raffel16}, which is a superset of Slakh2100 that we use. Orchestration by Arranger is a note-by-note mapping process, where each note in the piano source is mapped to a multi-track target by assigning instruments under a classification framework.

\textbf{\textit{BL-Pno.}}: We introduce \emph{Poly-Dis} by \cite{wang2020learning} as our baseline for the piano cover generation task. This model is pre-trained on POP909 and is also based on style transfer. Specifically, it can generate piano cover for a multi-track source piece by reconceptualizing its chord progression using the texture from a piano reference. In our case, we provide Poly-Dis with the same piano reference as our model and extract the chord progression of the source music using the algorithm in \cite{jiang2019large}.

\textbf{\textit{BL-ReIns.}}: We introduce the model by \cite{hung2019musical} as our baseline for the re-instrumentation task. This model rearranges a source piece using the synthesized audio timbre feature from a reference piece as instrumentation style. We train this model on Slakh2100 using both the MIDI and the aligned audio that is synthesized using professional-level sample-based virtual instruments \cite{manilow2019cutting}.

Besides the baselines, we also introduce three variants of our model to analyze the impact of each key component. Specifically, \textbf{\textit{Q\&A-T}} uses only time function as query to rearrange a piece. \textbf{\textit{Q\&A-P}}, on the other hand, uses pitch function only. The final variant \textbf{\textit{Q\&A w/o Ins.}} uses both functions but is trained without instrument embedding. 

\subsubsection{Evaluation Details}
We invite participants to subjectively evaluate the rearrangement quality of all models through a double-blind online survey. Our survey consists of 15 rearrangement sets, each of which contains one original source piece followed by five rearrangement samples (four by our model variants and the rest by one of the baselines). Among the 15 sets, there are 5 for piano cover generation, orchestration, and re-instrumentation, respectively. The original piece $x$ is an 8-bar musical phrase randomly selected from the validation/test set of either POP909 or Slakh2100 depending on the task. The reference piece $y$ is sampled from the other dataset following Equation \eqref{search}, where we set $\alpha=0.2$. For re-instrumentation, both $x$ and $y$ are in Slakh2100 but from different splits (validation and test sets, respectively). We use the same $y$ for each model that requires a reference piece for style transfer. 

In our survey, we request each participant to listen to 5 rearrangement sets and evaluate each sample. Both the set order and the sample order in each set are randomized. The evaluation is based on a 5-point scale from 1 (very low) to 5 (very high) for four criteria as follows:
\begin{itemize}
    \item \textbf{DOA}: The degree of arrangement. A low DOA refers to a note-by-note copy-paste from the original music, while a high DOA means the music is appropriately restructured to fit the new track system and instruments.
    \item \textbf{Creativity}: How creative the rearrangement is.
    \item \textbf{Naturalness}: How likely a human arranger creates it.
    \item \textbf{Musicality}: The overall musicality.
\end{itemize}

\subsubsection{Overall Rearrangement Performance}
\begin{figure}
 \centerline{
 \includegraphics[width=\columnwidth]{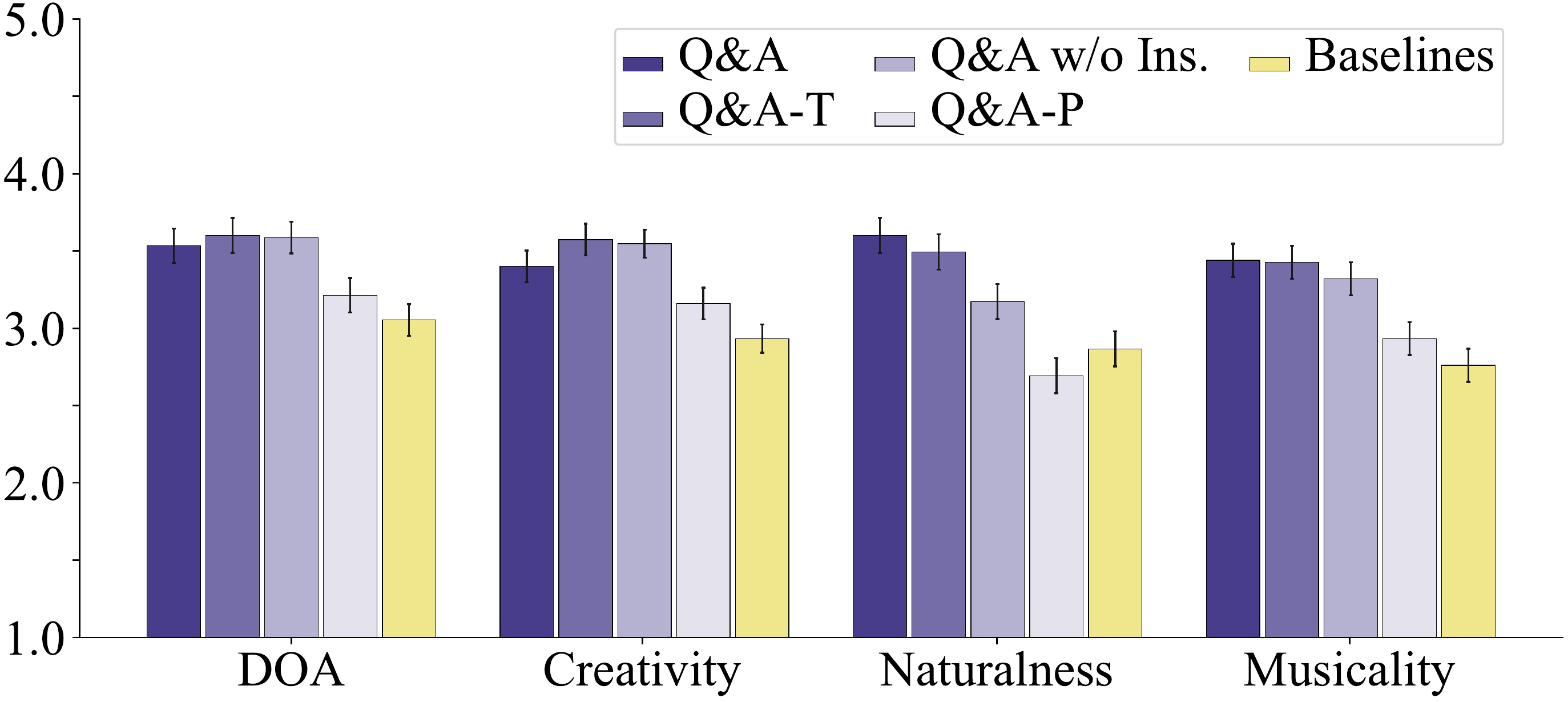}}
 \caption{Evaluation on overall rearrangement performance.}
 \label{overall_rating}
\end{figure}

A total of 26 participants (8 females and 18 males) with various musical backgrounds have completed our survey. We first show the statistical results for \emph{overall} rearrangement performance disregarding the specific tasks. As shown in Figure \ref{overall_rating}, the height of each bar represents the mean rating value and the error bar represents the standard error computed via within-subject (repeated-measures) ANOVA \cite{scheffe1999analysis}. Among our model variants, \textit{Q\&A-T} queries the mixture by time function only, and \textit{Q\&A w/o Ins.} has no instrument embedding. Both models essentially have fewer constraints during training and hence can produce more diverse results, which may explain the higher ratings on DOA and Creativity for both models. However, such results can also be less natural or musical. On the other hand, \textit{Q\&A-P} uses pitch function only and yields results inferior to other variants. This finding shows that pitch function alone is not sufficient to capture track structures in multi-track music. Indeed, pop music (at least in our datasets) is generally better characterized in grooves than in chords, as the latter can often fit in a few off-the-shelf template progressions. In terms of Naturalness and Musicality, our standard \textit{Q\&A} model makes a better balance and acquires significantly better results (p-value $p<0.01$) than all variants and the baselines ensemble.

\subsubsection{Task-Specific Performance}
\begin{figure}
 \centerline{
 \includegraphics[width=\columnwidth]{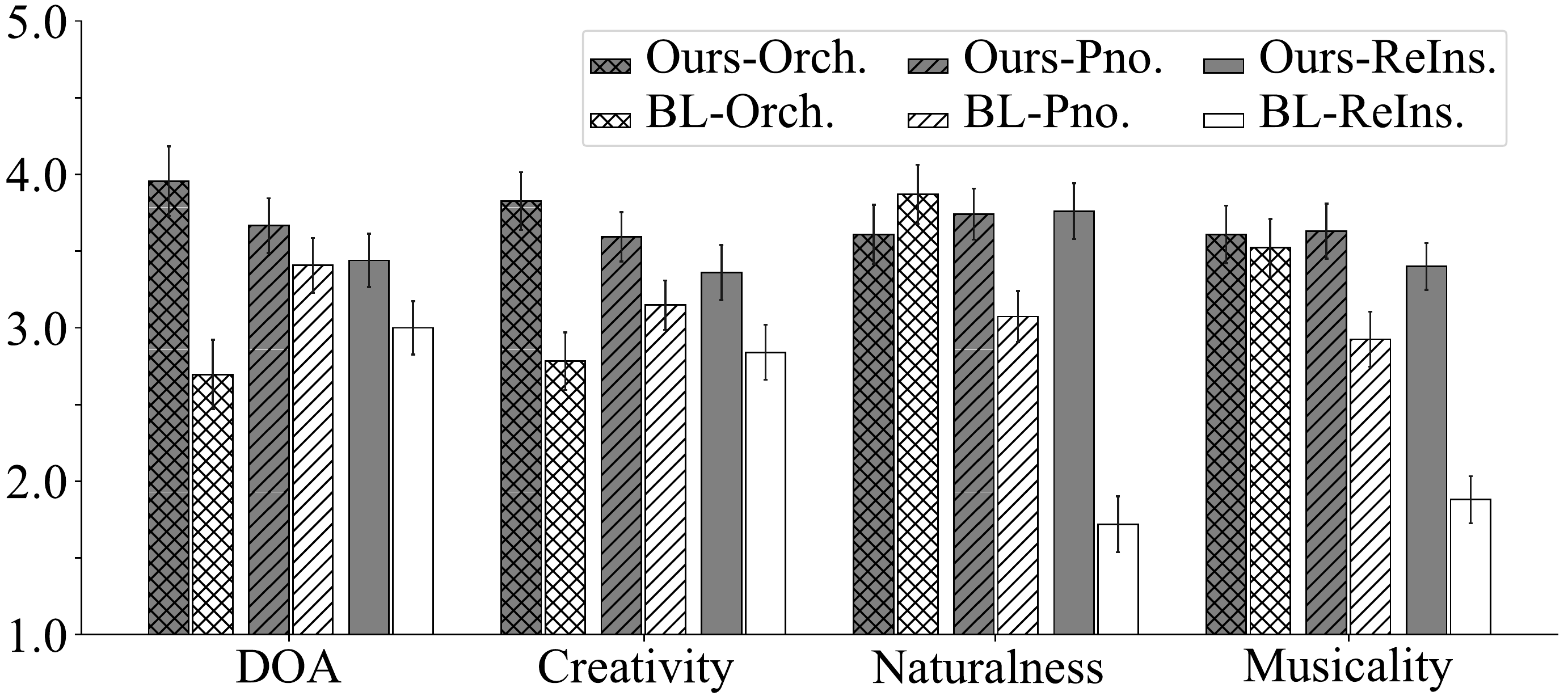}}
 \caption{Evaluation on task-specific rearrangement performance.}
 \label{task_specific}
\end{figure}

We are also interested in our model's performance on each concrete rearrangement task. As shown in Figure \ref{task_specific}, we show our model's \emph{task-specific} ratings (top in four variants) on the same set of criteria in comparison to corresponding baseline models. We notice that \textit{BL-Orch.} earns the highest rating for Naturalness in the orchestration task, which is not surprising because it adopts a note-by-note mapping strategy that virtually reproduces the original human-created music. Accompanied by this strategy is a lower degree of orchestration (DOA) and Creativity. On the other hand, our model demonstrates a more balanced and superior performance. It also outperforms \textit{BL-Orch.} in Musicality as it can introduce more diversified instruments and properly rearrange the source music with new texture and voicing. In particular, we report a significantly better performance (p-values $p<0.05$) of our model in Musicality than all baselines in all three tasks.

\subsection{Objective Evaluation on Voice Separation}\label{voice_sep}
One may wonder if the exceptional performance of our model in creativity and musicality sacrifices the faithfulness to the original music. To this end, we conduct an additional experiment on the task of voice separation and compare our model with note-by-note decision models --- a BiLSTM and a Transformer encoder tailored for this task in \cite{dong2021towards}. Specifically, voice separation is a special case of orchestration, where we only aim to separate a mixture into individual voicing tracks without any creative factor. Note that, in such a case, note-by-note classification methods have a natural advantage over representation learning based methods because the latter gives less importance to accurate control on low-level tokens. Still, the faithfulness of our model can be validated if it can also tackle this problem.



In voice separation, since the goal is to separate individual tracks, the ground-truth track functions cannot be the model input. Hence we introduce a new variant \textit{Q\&A-V}, which applies an additional GRU decoder to infer function representations $z^{f(x)}_{1:N}$ from mixture representation $z_\mathrm{mix}^x$, and then generate each track with inferred track functions. In our case, $N=4$ is preset and the inference process is conducted from high voice to low voice autoregressively. We load the rest part of the model with pre-trained parameters from standard \textit{Q\&A} and fine-tune the whole model on string quartets in MusicNet \cite{thickstun2016learning} and Bach chorales in Music21 \cite{cuthbert2010music21}, respectively. We process the data into 8-beat segments irrespective of time signature. At test time, if a certain note in the mixture is not recalled by our model, we look for its nearest-neighbour note that is generated and assign its voice. If two note assignments form polyphonic voice, we then re-assign the note with least added distance to its second-nearest voice, which is a simple greedy-based rule. As both datasets are tiny and prone to unbalanced train-test split, we evaluate our model by 10-fold cross validation. 


We show the test results (percentage accuracy) in Table \ref{voice_separation_results}. Compared to the baselines, our \textit{Q\&A-V} model yields generally comparable results, although with a noticeable gap on Bach chorales. Specifically, Bach chorales come with very regular and transparent counterpoints, which are a good fit for note-by-note classification frameworks to separate voices. On the other hand, string quartets have much more complex and even overlapped voices that are harder to separate. For this case, our model yields highly competitive performance in general. When entry hints are provided, our model achieves the best with a good margin to both baselines. 


\begin{table}
  \centering
    \begin{tabular}{lcc}
    \toprule
    \textbf{Model} & \multicolumn{1}{l}{\textbf{Chorales}} & \multicolumn{1}{l}{\textbf{Quartets}} \\
    \midrule
    Q\&A-V & 94.84$^\dag$ & 73.47$^\dag$ \\
    Transformer & 96.81 & 58.86 \\
    BiLSTM & \textbf{97.13} & \textbf{74.38} \\
    \midrule
    Q\&A-V (+ entry hints) & 95.11$^\dag$   & \textbf{78.71}$^\dag$ \\
    Transformer (+ entry hints) & 93.81 & 56.72 \\
    BiLSTM (+ entry hints) & \textbf{97.39} & 71.51 \\
    \bottomrule
    \end{tabular}%
    \caption{Objective evaluation on voice separation comparing to note-by-note architectures. We use $^\dag$ to denote test results under 10-fold cross validation. Baseline results are from \protect\cite{dong2021towards}.}
  \label{voice_separation_results}%
\end{table}%

\section{Conclusion}
In conclusion, we contribute Q\&A, a novel query-based framework for multi-track music rearrangement. The main novelty lies first in our application of a style transfer methodology to interpret the general rearrangement problem. By defining and utilizing track functions, we effectively capture the texture and voicing structure of multi-track music as composition style. Under a self-supervised query system, the number of tracks and instruments to rearrange a piece is virtually unconstrained. Q\&A serves as a unified solution for piano cover generation, orchestration, re-instrumentation, and voice separation. Extensive experiments prove that it can both creatively rearrange a piece and faithfully conserve the essential structures. We believe that our contributions will inspire further advancements in computer music research, opening doors to broader possibilities for universal music co-creation.

\bibliographystyle{named}
\bibliography{ijcai23}

\end{document}